# Selecting wavelengths for least squares range estimation

Assad Akhlaq, Robby McKilliam, Ramanan Subramanian and André Pollok

*Abstract*—We consider the problem of estimating the distance, or range, between two locations by measuring the phase of multiple sinusoidal signals transmitted between the locations. Traditional estimators developed for optical interferometry include the beat wavelength and excess fractions methods. More recently, estimators based on the Chinese remainder theorem (CRT) and least squares have appeared. Recent research suggests the least squares estimator to be most accurate in many cases. The accuracy of all of these range estimators depends upon the wavelengths chosen. This leads to the problem of selecting wavelengths that maximise accuracy. Procedures for selecting wavelengths for the beat wavelength and excess fractions methods have previously been described, but procedures for the CRT and least squares estimators are yet to be developed. In this paper we develop an algorithm to automatically select wavelengths for use with the least square range estimator. The algorithm minimises an optimisation criterion connected with the mean square error. Interesting properties of a particular class of *lattices* simplify the criterion allowing minimisation by depth first search. Monte-Carlo simulations indicate that wavelengths that minimise the criterion can result is considerably more accurate range estimates than wavelengths selected by ad hoc means.

## I. Introduction

Range (or distance) estimation is an important component in technologies such as electronic surveying [1, 2], global positioning [3, 4], and ranging cameras [5, 6]. Common methods of range estimation are based upon received signal strength [7, 8], time of flight (or time of arrival) [9, 10], and phase of arrival [11, 12]. This paper focuses on the phase of arrival method which provides the most accurate range estimates in many applications. Phase of arrival has become the technique of choice in modern high precision surveying and global positioning [13–16].

A difficulty with phase of arrival is that only the principal component of the phase can be observed. This limits the range that can be unambiguously estimated. This is sometimes referred to as the problem of *phase ambiguity* and it is related to what has been called the *notorious wrapping problem* in the circular statistics and meteorology literature [17]. One approach to address this problem is to utilise signals of multiple different wavelengths and observe the phase at each. The range can then be measured within an interval of length equal to the least common multiple of the wavelengths. Range estimators from such observations have been studied by numerous authors. Techniques include the beat wavelength method of Towers et al. [18, 19], the method of excess fractions [20–23], and methods based on the Chinese Remainder Theorem (CRT) [24–31]. Least squares/maximum likelihood and maximum a posteriori estimators of range have been studied by Teunissen [4], Hassibi and Boyd [32], and more recently by Li et al. [33] and Akhlaq et al. [34]. A key realisation is that the least squares estimator can be efficiently computed by solving a well known integer programming problem, that of computing a *closest point* in a *lattice* [35]. Teunissen [4] appears to have been the first to have realised this connection.

The accuracy of all of these range estimators depends upon the wavelengths chosen. This naturally leads to the problem of selecting wavelengths that maximise accuracy. This selection procedure is typically subject to practical constraints such minimum and maximum wavelength (i.e. bandwidth constraints) and constraints on the maximum identifiable range. The relationship between wavelengths and range estimation accuracy is nontrivial and this complicates wavelength selection procedures. Procedures have been described for the beat wavelength method [18] and for the method of excess fractions [21].

In this paper we develop an algorithm to automatically select wavelengths for use with the least square range estimator from [34]. We devise an optimisation criterion connected with the mean square error under constraints on the minimum and maximum wavelength and on the identifiable range. The optimisation criterion is developed using the interesting properties of a particular class of *lattices*, a structure common in algebraic and computational number theory [36–39]. These properties lead to simple and sufficiently accurate approximations for the mean square range error in terms of the wavelengths. The resulting constrained optimisation problem is simple enough to be minimised by a depth first search. Monte-Carlo simulations indicate that wavelengths that minimise this criterion can result is considerably more accurate range estimates than wavelengths selected by ad hoc means.

The paper is organised as follows. Section II introduces the signal model for range estimation from multiple phase observations and describes the least squares range estimator. Section III describes required properties from the theory of lattices [36–39]. The interesting properties of a particular class of lattices constructed by intersection with and projection onto a subspace are described. These properties are used to develop simple and sufficiently accurate approximations for the mean square error of the least squares range estimator in

The authors are at the Institute for Telecommunications Research at the School of Information Technology and Mathematical Sciences, The University of South Australia, SA, 5095. Supported by Australian Research Council Linkage Project LP130100514. Email: assad.akhlaq@mymail.unisa.edu.au, {robby.mckilliam, ramanan.subramanian, andre.pollok}@unisa.edu.au.

Section IV. Section V uses these approximations to design an optimisation criterion related to the mean square error and describes an algorithm to compute wavelengths that minimise the criterion. The algorithm is based on depth first search and can take a long time when the number of wavelength is not small. For this case we describe two methods that reduce the search time at the expense of not guaranteeing that the true minimising wavelengths are found. The results of Monte-Carlo simulations are presented in Section VI. The simulations indicate that wavelengths that minimise (or approximately minimise) the criterion can result is considerably more accurate range estimates than wavelengths selected by ad hoc means. The simulations corroborate with existing empirical evidence suggesting that the least squares range estimator is often more accurate than other estimators [34, 40].

## II. THE LEAST SQUARES RANGE ESTIMATOR

Suppose that a transmitter sends a signal of the form

$$x(t) = \sin(2\pi f t + 2\pi \phi) \qquad (1)$$

of known phase $\phi$ and frequency $f$ in Hertz. The signal is assumed to propagate by line of sight to a receiver resulting in the signal

$$y(t) = \alpha x(t - r_0/c) + \omega(t) = \alpha \sin(2\pi f t + 2\pi \theta) + \omega(t) \quad (2)$$

where $r_0$ is the distance (or range) in meters between receiver and transmitter, $c$ is the speed at which the signal propagates in meters per second, $\alpha$ is the amplitude of the received signal, $\omega(t)$ represents noise,

$$\theta = \phi - \frac{f}{c} r_0 = \phi - \frac{r_0}{\lambda} \qquad (3)$$

is the phase of the received signal, and $\lambda = c/f$ is the wavelength. Alternatively, the transmitter and receiver could be in the same location and the receiver obtains the signal after being reflected off a target. In this case, the range of the target would be $r_0/2$. The receiver is assumed to be *synchronised* by which it is meant that the phase $\phi$ and frequency $f$ are known to the receiver.

Our aim is to estimate $r_0$ from the signal $y(t)$. To do this we first calculate an estimate $\hat{\theta}$ of the principal component of the phase $\theta$. In optical ranging applications the phase estimate might be given by an interferometer. In sonar or radio frequency ranging applications an estimate might be obtained from samples of the signal $y(t)$ after demodulation. Whatever the method of phase estimation, the distance $r_0$ between receiver and transmitter is related to $\hat{\theta}$ by the phase difference

$$Y = \langle \phi - \hat{\theta} \rangle = \langle r_0/\lambda + \Phi \rangle, \qquad (4)$$

where $\Phi$ represents phase noise and $\langle x \rangle = x - \lceil x \rfloor$. The notation $\lceil x \rfloor$ denotes the closest integer to $x$ with half integers rounded up. For all integers $k$ we have

$$Y = \langle r_0/\lambda + \Phi \rangle = \langle (r_0 + k\lambda)/\lambda + \Phi \rangle \qquad (5)$$

and so ranges $r_0$ and $r_0 + k\lambda$ result in the same phase difference. For this reason the range is identifiable from the phase only if we assume $r_0$ to lie in some interval of length $\lambda$. A natural choice is the interval $[0, \lambda)$. This poses a problem if the range $r_0$ is larger than the wavelength $\lambda$.

A common approach to address this problem is to transmit multiple signals at multiple different frequencies and observe the phase at each. In this approach, $N$ phase estimates $\hat{\theta}_1, \ldots, \hat{\theta}_N$ and $N$ phase differences

$$Y_n = \langle \phi - \hat{\theta}_n \rangle = \langle r_0/\lambda_n + \Phi_n \rangle, \qquad n = 1, \ldots, N \quad (6)$$

are computed, where $\lambda_n = c/f_n$ is the wavelength of the $n$th signal and $\Phi_1, \ldots, \Phi_N$ represent phase noise. Let

$$P = \text{lcm}(\lambda_1, \ldots, \lambda_N)$$

be the least common multiple of the wavelengths. The least common multiple is the smallest positive integer such that $P/\lambda_1, \ldots, P/\lambda_N$ are all integers. Observe that the ranges $r_0$ and $r_0 + kP$ for any integer $k$ result in the same phase differences $Y_1, \ldots, Y_N$ and so $r_0$ can be uniquely identified only within an interval of length $P$. A natural choice is the interval $[0, P)$. The least common multiple $P$ is typically much larger than any individual wavelength and so the identifiable range can be considerably enlarged by the use of multiple wavelengths. If $\lambda_n/\lambda_m$ is irrational for any $n$ and $m$ then the least common multiple $P$ does not exist. In this paper we assume this is not the case and that a finite least common multiple $P$ does exist. This is a common assumption in the literature and in practice.

To motivate our wavelength selection procedure we make the assumption that the phase noise $\Phi_1, \ldots, \Phi_n$ are zero mean, independent and identically distributed (i.i.d.) wrapped normal random variables [41, p. 50][39, p. 76][17, p. 47]. In this case,

$$\Phi_n = \langle \epsilon_n \rangle \qquad n = 1, \ldots, N$$

where $\epsilon_1, \ldots, \epsilon_N$ are independent and identically distributed normal random variables with zero mean and variance $\sigma^2$. Under this assumption, the least squares range estimator from [34] is also the maximum likelihood estimator. Observe that

$$Y = \langle r_0/\lambda_n + \Phi_n \rangle = \langle r_0/\lambda_n + \langle \epsilon_n \rangle \rangle = \langle r_0/\lambda_n + \epsilon_n \rangle$$

and that the phase differences can be written in the form

$$Y_n = \langle r_0/\lambda_n + \epsilon_n \rangle = r_0/\lambda_n + \epsilon_n + \zeta_n$$

where the integers

$$\zeta_n = -\lceil r_0/\lambda_n + \epsilon_n \rfloor \qquad n = 1, \ldots, N$$

are called *wrapping variables*. The wrapping variables are related to the number of whole wavelengths that occur over the range $r_0$ between the transmitter and the receiver. Writing in column vector form

$$\mathbf{y} = r_0 \mathbf{w} + \boldsymbol{\epsilon} + \boldsymbol{\zeta} \qquad (7)$$

where the column vectors

$$\mathbf{y} = \begin{pmatrix} Y_1 \\ \vdots \\ Y_N \end{pmatrix} \quad \boldsymbol{\zeta} = \begin{pmatrix} \zeta_1 \\ \vdots \\ \zeta_N \end{pmatrix} \quad \mathbf{w} = \begin{pmatrix} \frac{1}{\lambda_1} \\ \vdots \\ \frac{1}{\lambda_N} \end{pmatrix} \quad \boldsymbol{\epsilon} = \begin{pmatrix} \epsilon_1 \\ \vdots \\ \epsilon_N \end{pmatrix}.$$

The $n$th element of the vector $\mathbf{w}$ is the reciprocal of the $n$th wavelength, that is, $w_n = 1/\lambda_n$. Observe that $P$ is the smallest positive number such that the vector

$$\mathbf{v} = P\mathbf{w} = (P/\lambda_1, \ldots, P/\lambda_N) \in \mathbb{Z}^N,$$

that is, such that the elements of $\mathbf{v} = P\mathbf{w}$ are all integers. Equivalently, $P$ is the unique positive real number such that the elements of $\mathbf{v}$ are jointly relatively prime, that is, such that

$$\gcd(v_1, \ldots, v_N) = \gcd(P/\lambda_1, \ldots, P/\lambda_N) = 1.$$

Many range estimators, such as the least squares estimator and those estimators based on the CRT operate in two stages. In the first stage, an estimate $\hat{\zeta}$ of the wrapping variables $\zeta$ is made. Given $\hat{\zeta}$, an estimate of the range $r_0$ is typically given by linear regression, that is,

$$\hat{r} = \frac{(\mathbf{y} - \hat{\zeta})'\mathbf{w}}{\mathbf{w}'\mathbf{w}} \qquad (8)$$

where superscript $'$ indicates the vector or matrix transpose. For any integer $k$, the ranges $r_0$ and $r_0 + kP$ are equivalent and so range estimates $\hat{r}$ and $\hat{r} + kP$ for any integer $k$ are equivalent. It follows that estimates $\hat{\zeta}$ and $\hat{\zeta} + kP\mathbf{w}$ of the wrapping variables are equivalent, because

$$\frac{(\mathbf{y} - \hat{\zeta} + kP\mathbf{w})'\mathbf{w}}{\mathbf{w}'\mathbf{w}} = \hat{r} + kP.$$

For this reason, the estimated wrapping variables $\hat{\zeta}$ are to be considered error free (or correct), if $\hat{\zeta} = \zeta + kP\mathbf{w}$ for some integer $k$. Because $P$ is the smallest positive integer such that $\mathbf{v} = P\mathbf{w} \in \mathbb{Z}^N$ this occurs if and only if $\mathbf{Q}\hat{\zeta} = \mathbf{Q}\zeta$ where

$$\mathbf{Q} = \mathbf{I} - \frac{\mathbf{w}\mathbf{w}'}{\mathbf{w}'\mathbf{w}} = \mathbf{I} - \frac{\mathbf{v}\mathbf{v}'}{\mathbf{v}'\mathbf{v}} \qquad (9)$$

is the $N \times N$ orthogonal projection matrix onto the $N-1$ dimensional subspace orthogonal to $\mathbf{w}$ and $\mathbf{I}$ is the $N \times N$ identity matrix. In what follows, estimates $\hat{\zeta}$ of the wrapping variables $\zeta$ are said to be *correct* if $\mathbf{Q}\hat{\zeta} = \mathbf{Q}\zeta$.

It is shown in [34], that the least squares estimator $\hat{\zeta} \in \mathbb{Z}^N$ of the wrapping variables minimises the quadratic form

$$\|\mathbf{Q}\mathbf{y} - \mathbf{Q}\mathbf{z}\|^2 \qquad \text{over } \mathbf{z} \in \mathbb{Z}^N, \qquad (10)$$

where $\|\cdot\|$ indicates the Euclidean norm of a vector. Given $\hat{\zeta}$, the least square range estimator $\hat{r}$ is then given by (8). It is shown in [34] how the quadratic form (10) can be minimised over $\mathbb{Z}^N$ by computing a closest point in a *lattice*. We will use the properties of this lattice to develop our wavelength selection procedure. We first require some concepts from lattice theory.

## III. LATTICE THEORY

Let $\mathbf{b}_1, \ldots, \mathbf{b}_n$ be linearly independent vectors from $m$-dimensional Euclidean space $\mathbb{R}^m$ with $m \geq n$. The set of vectors

$$\Lambda = \{u_1\mathbf{b}_1 + \cdots + u_n\mathbf{b}_n \; ; \; u_1, \ldots, u_n \in \mathbb{Z}\}$$

is called an $n$-dimensional *lattice*. The elements of $\Lambda$ are called *lattice points* or *lattice vectors*. The vectors $\mathbf{b}_1, \ldots, \mathbf{b}_n$ form a *basis* for the lattice $\Lambda$. We can equivalently write

$$\Lambda = \{\mathbf{B}\mathbf{u} \; ; \; \mathbf{u} \in \mathbb{Z}^n\}$$

where $\mathbf{B}$ is the $m \times n$ matrix with columns $\mathbf{b}_1, \ldots, \mathbf{b}_n$. The matrix $\mathbf{B}$ is called a *basis* or *generator* for $\Lambda$. The set of integers $\mathbb{Z}^n$ is called the *integer lattice* with the $n \times n$ identity matrix $\mathbf{I}$ as a basis. When $m > n$ the lattice points lie in the $n$-dimensional subspace of $\mathbb{R}^m$ spanned by $\mathbf{b}_1, \ldots, \mathbf{b}_n$. The parallelepiped formed by basis vectors $\mathbf{b}_1, \ldots, \mathbf{b}_n$ is called a *fundamental parallelepiped* of the lattice $\Lambda$. A fundamental parallelepiped has $n$-dimensional volume $\sqrt{\det \mathbf{B}'\mathbf{B}}$ where $\det \mathbf{B}'\mathbf{B}$ is the determinant of the $n \times n$ matrix $\mathbf{B}'\mathbf{B}$. This quantity is also called the *determinant* of the lattice and is denoted by $\det \Lambda$.

Let $\Lambda$ be an $n$-dimensional lattice and let $H$ be the $n$-dimensional subspace spanned by its lattice points. The *dual lattice* of $\Lambda$, denoted $\Lambda^*$, contains those points from $H$ that have integer inner product with all points from $\Lambda$, that is,

$$\Lambda^* = \{\mathbf{x} \in H \; ; \; \mathbf{x}'\mathbf{y} \in \mathbb{Z} \text{ for all } \mathbf{y} \in \Lambda\}.$$

The determinant of a lattice and its dual are reciprocals, that is, $\det \Lambda = (\det \Lambda^*)^{-1}$ [36, p. 10]. A lattice and its dual have interesting properties when intersected with or projected onto a subspace.

**Proposition 1.** *Let $\Lambda \subset \mathbb{R}^n$ be an $n$ dimensional lattice, and let $H$ be an $n - k$ dimensional subspace of $\mathbb{R}^n$. Let $H^\perp$ be the $k$ dimensional space orthogonal to $H$ and let $p$ be the orthogonal projection onto $H$. The set $\Lambda \cap H$ is an $n - k$ dimensional lattice if and only if $\Lambda \cap H^\perp$ is a $k$ dimensional lattice. Moreover, if $\Lambda \cap H$ is an $n - k$ dimensional lattice then:*

1) *The dual of $\Lambda \cap H$ is the orthogonal projection of $\Lambda^*$ onto $H$, that is, $(\Lambda \cap H)^* = p(\Lambda^*)$.*
2) *The determinants of $\Lambda$, $\Lambda \cap H$ and $\Lambda^* \cap H^\perp$ are related by $\det(\Lambda) \det(\Lambda^* \cap H^\perp) = \det(\Lambda \cap H)$.*

*Proof:* Proposition 1.3.4 and Corollary 1.3.5 of [37]. ∎

For the purpose of developing our wavelength optimisation criterion we will be particularly interested in Proposition 1 when $\Lambda$ is the integer lattice $\mathbb{Z}^n$ and $k = 1$. We state this special case in the following corollary. The corollary makes use of the fact that the integer lattice is *self-dual*, that is, $\mathbb{Z}^n = (\mathbb{Z}^n)^*$.

**Corollary 1.** *Let $\mathbf{v} \in \mathbb{Z}^n$ be a vector of jointly relatively prime integers, let $H$ be the $n-1$ dimensional subspace orthogonal to $\mathbf{v}$, and let*

$$\mathbf{Q} = \mathbf{I} - \frac{\mathbf{v}\mathbf{v}'}{\mathbf{v}'\mathbf{v}} = \mathbf{I} - \frac{\mathbf{v}\mathbf{v}'}{\|\mathbf{v}\|^2}$$

*be the $n \times n$ orthogonal projection matrix onto $H$. The set of vectors $\mathbb{Z}^n \cap H$ is an $n-1$ dimensional lattice with determinant*

$$\det(\mathbb{Z}^n \cap H) = \|\mathbf{v}\|$$

*and dual lattice*

$$(\mathbb{Z}^n \cap H)^* = \{\mathbf{Q}\mathbf{z} \; ; \; \mathbf{z} \in \mathbb{Z}^n\}.$$

The (closed) *Voronoi cell*, denoted $\operatorname{Vor}\Lambda$, of an $n$-dimensional lattice $\Lambda$ in $\mathbb{R}^m$ is the subset of $\mathbb{R}^m$ containing all points nearer or of equal distance (here with respect to the Euclidean norm) to the lattice point at the origin than to any other lattice point. If the lattice is full rank so that $n = m$ then the volume of the Voronoi cell is equal to the volume of a fundamental parallelepiped, that is, $\det \Lambda$. Otherwise, if $m > n$ the Voronoi cell is unbounded in those directions orthogonal to the subspace spanned by the basis vectors $\mathbf{b}_1, \ldots, \mathbf{b}_n$. Specifically, if $\mathbf{x}$ is contained in this orthogonal subspace, then $\mathbf{y} \in \operatorname{Vor}\Lambda$ if and only if $\mathbf{y} + s\mathbf{x} \in \operatorname{Vor}\Lambda$ for all $s \in \mathbb{R}$. In this case, the intersection of the Voronoi cell with the subspace spanned by $\mathbf{b}_1, \ldots, \mathbf{b}_n$ has $n$-dimensional volume equal to $\det \Lambda$.

A *short vector* in a lattice $\Lambda$ is a lattice point of minimum nonzero Euclidean length, that is, a lattice point of length

$$d_{\min} = \min_{\mathbf{x} \in \Lambda \setminus \{\mathbf{0}\}} \|\mathbf{x}\|^2.$$

The length $d_{\min}$ of a short vector is the smallest distance between any two lattice points. The *inradius* or *packing radius* $\rho = d_{\min}/2$ is the length of a point on the boundary of the Voronoi cell that is closest to the origin (Figure 1). Equivalently, the inradius is the radius of the largest sphere that fits inside the Voronoi cell. It is also the radius of the largest sphere that can be centered at each lattice point such that no two spheres intersect. Such an arrangement of spheres is called a *sphere packing* (Figure 1).

Of interest to us is the probability that an $m$-variate normal random variable with i.i.d. components having zero mean and variance $\sigma^2$ lies inside the Voronoi cell. We denote this probability by

$$P(\Lambda, \sigma^2) = \frac{1}{\sigma^m \sqrt{(2\pi)^m}} \int_{\operatorname{Vor}\Lambda} e^{-\|\mathbf{x}\|^2/2\sigma^2} d\mathbf{x}. \quad (11)$$

This probability can be upper bounded by the probability that an $n$-variate normal random variable lies within a sphere of $n$-volume equal to the determinant of the lattice $\det \Lambda$ [32, Sec. IV.C], i.e.,

$$P(\Lambda, \sigma^2) \leq F_n\left(\frac{\Gamma(n/2+1)^{2/n}(\det\Lambda)^{2/n}}{\pi\sigma^2}\right) \quad (12)$$

where $F_n$ is the chi-square cumulative distribution function with $n$ degrees of freedom and $\Gamma$ is the gamma function. This upper bound will be involved in the construction of our wavelength optimisation criterion in Section V. The probability $P(\Lambda, \sigma^2)$ can be lower bounded by the probability that an $n$-variate normal random variable lies within a sphere of radius equal to the inradius $\rho$ of the lattice, i.e.,

$$P(\Lambda, \sigma^2) \geq F_n(\rho/\sigma^2). \quad (13)$$

It may be possible to build an alternative wavelength optimisation criterion using this lower bound rather than (12). However, the relationship between the wavelengths and the inradius $\rho$ is nontrivial and so we have not attempted this here.

Given a lattice $\Lambda$ in $\mathbb{R}^m$ and a vector $\mathbf{y} \in \mathbb{R}^m$, a problem of interest is to find a lattice point $\mathbf{x} \in \Lambda$ such that the squared Euclidean norm

$$\|\mathbf{y} - \mathbf{x}\|^2 = \sum_{i=1}^{m}(y_i - x_i)^2$$

is minimised. This is called the *closest lattice point problem* (or *closest vector problem*) and a solution is called a *closest lattice point* (or simply *closest point*) to $\mathbf{y}$ [35, 42, 43]. The problem has found numerous applications in computer science, engineering, and statistics [44–52]. The closest lattice point problem and the Voronoi cell are related in that $\mathbf{x} \in \Lambda$ is a closest lattice point to $\mathbf{y}$ if and only if $\mathbf{y} - \mathbf{x} \in \operatorname{Vor}\Lambda$. The closest lattice point is not necessarily unique, that is, there can be multiple lattice points that minimise $\|\mathbf{y} - \mathbf{x}\|^2$. This occurs precisely when $\mathbf{y} - \mathbf{x}$ lies on the boundary of $\operatorname{Vor}\Lambda$. If $\mathbf{y} - \mathbf{x}$ is contained strictly in the interior of $\operatorname{Vor}\Lambda$, then $\mathbf{x} \in \Lambda$ is the unique closest lattice point to $\mathbf{y}$. In particular, if $\mathbf{y}$ itself is in the interior of $\operatorname{Vor}\Lambda$, then the unique closest lattice point to $\mathbf{y}$ is the origin $\mathbf{0}$.

Recall from (10) that the least squares range estimator first computes an estimate $\hat{\boldsymbol{\zeta}} \in \mathbb{Z}^N$ of the wrapping variables by minimising the quadratic form $\|\mathbf{Qy} - \mathbf{Qz}\|^2$ with respect to $\mathbf{z} \in \mathbb{Z}^N$. The $N \times N$ matrix $\mathbf{Q}$ is the orthogonal projection into the $N-1$ dimensional subspace orthogonal to the vector $\mathbf{w}$ containing the reciprocals of the wavelengths (9). Let $H$ denote this subspace. The elements in the vector $\mathbf{v} = P\mathbf{w}$ are jointly relatively prime and so, by Corollary 1, the set $\Lambda = \mathbb{Z}^N \cap H$ is an $N-1$ dimensional lattice with determinant $\det \Lambda = \|\mathbf{v}\|$ and dual lattice $\Lambda^* = \{\mathbf{Qz} \; ; \; \mathbf{z} \in \mathbb{Z}^N\}$. We see that the problem of minimising the quadratic form (10) is precisely that of finding a closest lattice point to $\mathbf{Qy}$ in the lattice $\Lambda^*$.

This connection between the least squares range estimator and the closest lattice point problem appears to have been first realised by Teunissen [4]. The notation we use here and the connection with Corollary 1 first appeared in [34]. In the next section we will show that the least squares estimator $\hat{\boldsymbol{\zeta}}$ of the wrapping variables is correct when the noise $\boldsymbol{\epsilon}$ is contained within the Voronoi cell of the lattice $\Lambda^*$. This fact has been realised by numerous authors including Hassibi and Boyd [32] who relate it to what they call the problem of *verification*. More recently this has been utilised by Li et al. [33] and Akhlaq et al. [40] for studying the accuracy of the least squares range estimator.

The existing literature typically makes use of either the lower bound (13) based on the inradius $\rho$ of the lattice $\Lambda^*$ or the upper bound (12) based on the determinant $\det \Lambda^*$. The relationship between the wavelengths and the inradius is non trivial. So far in the literature $\det \Lambda^*$ has been computed by first finding a basis matrix $\mathbf{B}$ for the lattice $\Lambda^*$ and then computing the determinant directly as $\det \Lambda^* = \sqrt{\det \mathbf{B}'\mathbf{B}}$. The relationship between the wavelengths and the basis $\mathbf{B}$ is nontrivial [34] and the determinant of the $N \times N$ matrix $\mathbf{B}'\mathbf{B}$ is also not given by a simple expression when $N$ is not small. For these reasons it at first appears that the relationship between the wavelengths and the determinant $\det \Lambda^*$ is non trivial.

A key realisation we make in this paper is that $\det \Lambda^*$ is related to the wavelengths in a simple way by Corollary 1.

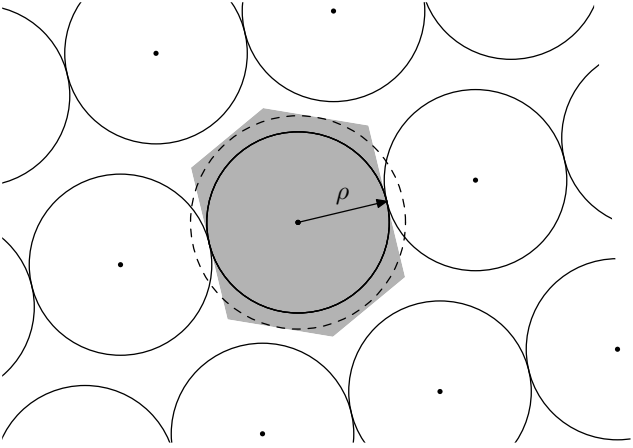

Fig. 1. The inradius $\rho = d_{\min}/2$ of the 2-dimensional lattice with basis $\mathbf{b}_1 = [3, 0.72]'$, $\mathbf{b}_2 = [0.6, 3.6]'$. The dots are the lattice points. The origin $\mathbf{0}$ is the lattice point in the center of the figure. This lattice has two short vectors. The shaded region shows the Voronoi cell of the lattice. The solid circles exhibit a sphere packing. The dashed circle has area (2-volume) equal to that of the Voronoi cell. The sphere with volume equal to the Voronoi cell is used in the upper bound (12).

This corollary and the fact that $\det \Lambda^* = (\det \Lambda)^{-1}$ shows that $\det \Lambda^*$ takes the simple form

$$\det \Lambda^* = \frac{1}{\|\mathbf{v}\|} = \frac{1}{\|P\mathbf{w}\|} = \frac{1}{P\sqrt{\sum_{i=1}^N \lambda_i^{-2}}}.$$

Combining this simple expression with the upper bound (12) will lead to a simple and sufficiently accurate approximation of the probability that the least square estimator of the unwrapping variables is correct, that is, an approximation of the probability that $\mathbf{Q}\hat{\zeta} = \mathbf{Q}\zeta$. It is this simple approximation that leads to our optimisation criterion for selecting wavelengths.

## IV. Approximating range error

In Section V we will describe a procedure for selecting favourable sets of wavelengths for the least squares range estimator. To do so, we first require approximations for the error of the range estimator in terms of the wavelengths $\lambda_1, \dots, \lambda_N$. We consider two approximations. The first approximates the error in the case that the least squares estimator of the wrapping variables $\hat{\zeta}$ is correct. The second uses (12) to upper bound the probability that $\hat{\zeta}$ is correct.

When the wrapping variables are correct $\mathbf{Q}\hat{\zeta} = \mathbf{Q}\zeta$ or equivalently $\zeta = \hat{\zeta} + kP\mathbf{w}$ for some $k \in \mathbb{Z}$. In this case, the least squares estimator of the range takes the form (8),

$$\hat{r} = \frac{(\mathbf{y} - \hat{\zeta})'\mathbf{w}}{\mathbf{w}'\mathbf{w}} = \frac{(\mathbf{y} - \zeta + kP\mathbf{w})'\mathbf{w}}{\|\mathbf{w}\|^2}.$$

Substituting (7) for $\mathbf{y}$ we find that

$$\hat{r} = \frac{\boldsymbol{\epsilon}'\mathbf{w}}{\|\mathbf{w}\|^2} + r_0 + kP.$$

Recall from Section II that range estimates $\hat{r}$ and $\hat{r} + kP$ are considered equivalent for integers $k$. For this reason the error of the least squares range estimator corresponds with the term $\boldsymbol{\epsilon}'\mathbf{w}/\|\mathbf{w}\|^2$. Under our assumption that $\epsilon_1, \dots, \epsilon_N$ are i.i.d.

and normally distributed with zero mean and variance $\sigma^2$ this error is normally distributed with zero mean and variance

$$\operatorname{var} \frac{\boldsymbol{\epsilon}'\mathbf{w}}{\|\mathbf{w}\|^2} = \frac{\sigma^2}{\|\mathbf{w}\|^2} = \frac{\sigma^2}{\sum_{n=1}^N \lambda_n^{-2}}. \qquad (14)$$

The variance decreases as $\sum_{n=1}^N \lambda_n^{-2}$ increases. The variance (14) serves as an approximation of the mean square error of the least squares range estimator when the estimated wrapping variables $\hat{\zeta}$ are correct. The simulation results in Section VI suggest this approximation to be very close.

We now approximate the probability that the wrapping variables are correct, that is, we approximate the probability that $\mathbf{Q}\hat{\zeta} = \mathbf{Q}\zeta$. Our approximation is based upon the upper bound (12). Recall from (10) that $\hat{\zeta}$ minimises the quadratic form $\|\mathbf{Q}\mathbf{y} - \mathbf{Q}\mathbf{z}\|^2$ over $\mathbf{z} \in \mathbb{Z}^N$. It follows that $\mathbf{Q}\hat{\zeta}$ is a closest point in the lattice $\Lambda^* = \{\mathbf{Q}\mathbf{z} \; ; \; \mathbf{z} \in \mathbb{Z}^N\}$ to the point $\mathbf{Q}\mathbf{y}$. Equivalently,

$$\mathbf{Q}\mathbf{y} - \mathbf{Q}\hat{\zeta} \in \operatorname{Vor} \Lambda^*$$

from the definition of the Voronoi cell. Using (7),

$$\mathbf{Q}\mathbf{y} - \mathbf{Q}\hat{\zeta} = \mathbf{Q}(r_0\mathbf{w} + \boldsymbol{\epsilon} + \zeta) - \mathbf{Q}\hat{\zeta} = \mathbf{Q}\boldsymbol{\epsilon} - \mathbf{Q}(\hat{\zeta} - \zeta)$$

and so

$$\mathbf{Q}\boldsymbol{\epsilon} - \mathbf{Q}(\hat{\zeta} - \zeta) \in \operatorname{Vor} \Lambda^*.$$

We see that $\mathbf{Q}(\zeta - \hat{\zeta})$ is a closest lattice point to the projection of the noise variables $\mathbf{Q}\boldsymbol{\epsilon}$. If $\mathbf{Q}\boldsymbol{\epsilon}$ lies in the interior of $\operatorname{Vor} \Lambda^*$ then the unique closest lattice point is the origin $\mathbf{0}$, that is, $\mathbf{Q}(\zeta - \hat{\zeta}) = \mathbf{0}$ or equivalently $\mathbf{Q}\zeta = \mathbf{Q}\hat{\zeta}$. We have found that the least square estimator $\hat{\zeta}$ of the unwrapping variables is correct if the projection $\mathbf{Q}\boldsymbol{\epsilon}$ of the noise variables lies within the interior of the Voronoi cell. The estimator $\hat{\zeta}$ can similarly be shown to be incorrect if $\mathbf{Q}\boldsymbol{\epsilon} \notin \operatorname{Vor} \Lambda^*$. Because the boundary of the Voronoi cell has zero $n$-volume, it follows that the probability the unwrapping variables are correct is the same as the probability that $\mathbf{Q}\boldsymbol{\epsilon}$ lies in $\operatorname{Vor} \Lambda^*$

A further simplification can be made. The $N-1$ dimensional lattice $\Lambda^*$ lies in the subspace orthogonal to $\mathbf{w}$ and so $\operatorname{Vor} \Lambda^*$ is unbounded in the direction of $\mathbf{w}$. Specifically, $\mathbf{Q}\boldsymbol{\epsilon} \in \operatorname{Vor} \Lambda^*$ if and only if $\mathbf{Q}\boldsymbol{\epsilon} + s\mathbf{w} \in \operatorname{Vor} \Lambda^*$ for all $s \in \mathbb{R}$. Because $\boldsymbol{\epsilon} = \mathbf{Q}\boldsymbol{\epsilon} + s\mathbf{w}$ for some $s$, it follows that $\mathbf{Q}\boldsymbol{\epsilon} \in \operatorname{Vor} \Lambda^*$ if and only if $\mathbf{Q}\boldsymbol{\epsilon} + s\mathbf{w} = \boldsymbol{\epsilon} \in \operatorname{Vor} \Lambda^*$, that is,

$$\mathbf{Q}\boldsymbol{\epsilon} \in \operatorname{Vor} \Lambda^* \Leftrightarrow \boldsymbol{\epsilon} \in \operatorname{Vor} \Lambda^*.$$

Thus, the probability that the unwrapping variables are correct is the same as the probability that the noise $\boldsymbol{\epsilon}$ lies in $\operatorname{Vor} \Lambda^*$, that is, the same as the probability $P(\Lambda^*, \sigma^2)$ from (11). We approximate this probability by the upper bound (12),

$$P(\Lambda^*, \sigma^2) \leq F_{N-1}\left(\frac{\Gamma\left(\frac{N}{2} + \frac{1}{2}\right)^{2/(N-1)}}{\|\mathbf{v}\|^{2/(N-1)}\sigma^2\pi}\right) \qquad (15)$$

where we have used the simple expression $\det \Lambda^* = \|\mathbf{v}\|^{-1}$ derived from Corollary 1. This bound depends upon the wavelengths $\lambda_1, \dots, \lambda_N$ only through the term

$$\|\mathbf{v}\|^2 = \|P\mathbf{w}\|^2 = P^2 \sum_{n=1}^N \lambda_n^{-2} = \operatorname{lcm}^2(\lambda_1, \dots, \lambda_N) \sum_{n=1}^N \lambda_n^{-2}.$$

The bound increases as this term decreases.

This bound (15) for the probability of correct unwrapping is simpler than similar bounds in the literature that involve computing the determinant $\det \Lambda^* = \sqrt{\det \mathbf{B}'\mathbf{B}}$ directly [32, 33]. The simplicity of our bound is made possible by Corollary 1 leading to the simple expression $\det \Lambda^* = \|\mathbf{v}\|^{-1}$. This simplicity enables the wavelength selection procedure we describe in the next section.

## V. Selecting wavelengths for range estimation

In the previous section two approximations, (14) and (15), related to range error were developed. The first approximation (14) describes the variance of the range error when the wrapping variables are correct. To decrease this variance we should choose wavelengths such that

$$\frac{\sigma^2}{\sum_{n=1}^{N} \lambda_n^{-2}}$$

is small. The second approximation (15) upper bounds the probability that the wrapping variables are correct. To increase this bound we should choose wavelengths such that

$$P^2 \sum_{n=1}^{N} \lambda_n^{-2} = \mathrm{lcm}^2(\lambda_1, \dots, \lambda_N) \sum_{n=1}^{N} \lambda_n^{-2}$$

is small.

These are two competing objectives. To have both small estimator variance while simultaneously allowing large probability of correct unwrapping we propose to choose wavelengths that minimise an objective function of the form

$$L(\lambda_1, \dots, \lambda_N) = P^2 \sum_{n=1}^{N} \lambda_n^{-2} + \frac{\gamma}{\sum_{n=1}^{N} \lambda_n^{-2}} \qquad (16)$$

where $\gamma > 0$ weights the importance of the individual objectives and is free to be chosen. The weight $\gamma$ can be chosen to incorporate $\sigma^2$ if it is known. We have found that choosing

$$\gamma = \frac{N^2 r_{\max}^2}{\lambda_{\max}^2 \lambda_{\min}^2}$$

works well empirically. The quantities $r_{\max}$, $\lambda_{\min}$, and $\lambda_{\max}$ will be introduced shortly. This choice for $\gamma$ is used in the experiments performed in Section VI and has the convenient property of being independent of the noise variance $\sigma^2$. The choice is approximately the ratio of the minimum value of the two individual objectives. The motivation behind this being to approximately balance the importance given to both objectives.

We incorporate into this optimisation problem three practical constraints. First, we suppose that the wavelengths are all contained in an interval $[\lambda_{\min}, \lambda_{\max}]$. In practice the minimum and maximum allowable wavelengths $\lambda_{\min}$ and $\lambda_{\max}$ might be dictated by hardware constraints, such as antennae size, or properties of the medium through which the signal propagates. The second constraint is upon the maximum identifiable range. We suppose that the system must be capable of unambiguously estimating range on an interval of some prespecified length $r_{\max}$, that is, $P \geq r_{\max}$. For example $r_{\max}$ maybe a few meters for indoor applications, a few tens of meters for outdoor electronic surveying, and a few thousand kilometers for global positioning via satellite. Finally, we assume that one of the wavelengths, say $\lambda_1$, is fixed and known. We assume that $\lambda_1 = \lambda_{\max}$ in what follows. This constraint simplifies the optimisation problem and, since $\lambda_{\max}$ is free to be selected, results in only minor loss of generality.

Our optimisation problem is now to find wavelengths $\lambda_2, \dots, \lambda_N$ that minimise

$$L_1(\lambda_2, \dots, \lambda_N) = L(\lambda_{\max}, \lambda_2, \dots, \lambda_N)$$

subject to constraints

$$\lambda_{\min} \leq \lambda_n \leq \lambda_{\max} \qquad n = 2, \dots, N \qquad (17)$$
$$P = \mathrm{lcm}(\lambda_{\max}, \lambda_2, \dots, \lambda_N) \geq r_{\max}. \qquad (18)$$

These are referred to as the *bandwidth constraint* and the *range constraint* respectively. The least common multiple $P = \mathrm{lcm}(\lambda_{\max}, \lambda_2, \dots, \lambda_N)$ depends upon the wavelengths in a non trivial way. This optimisation problem is multivariate, nonlinear, and nonconvex with nonconvex constraints. It is not immediately obvious how a solution is to be found. We will show how this problem can be transformed into an equivalent problem involving $2(N-1)$ integer parameters. This equivalent problem can be solved by a depth first search.

A solution of the minimisation problem is such that the wavelengths $\lambda_1 = \lambda_{\max}$ and $\lambda_2, \dots, \lambda_{N-1}$ are rationally related, that is, $\lambda_n/\lambda_m$ is rational for all $n, m$. Otherwise, $P = \mathrm{lcm}(\lambda_{\max}, \lambda_2, \dots, \lambda_N) = \infty$ and $L_1$ will not be minimised. Thus, there exist positive integers $p_2, \dots, p_N$ and $q_2, \dots, q_N$ such that $\gcd(p_n, q_n) = 1$ and

$$\lambda_n = \frac{p_n}{q_n}\lambda_1 = \frac{p_n}{q_n}\lambda_{\max} \qquad n = 2, \dots, N. \qquad (19)$$

Now,

$$P = \mathrm{lcm}\left(\lambda_{\max}, \frac{p_2}{q_2}\lambda_{\max}, \dots, \frac{p_N}{q_N}\lambda_{\max}\right) = \lambda_{\max} Q \qquad (20)$$

where

$$Q = \mathrm{lcm}\left(1, \frac{p_2}{q_2}, \dots, \frac{p_N}{q_N}\right).$$

A simpler expression for $Q$ can be obtained. Let $\ell_1, \dots, \ell_N$ satisfy

$$\ell_1 = 1, \qquad \ell_n = \mathrm{lcm}\left(\ell_{n-1}, \frac{p_n}{q_n}\right) \qquad n = 2, \dots, N$$

and observe that $Q = \ell_N$. Because $\ell_1$ is an integer and $p_2$ and $q_2$ are relatively prime $\ell_2 = \mathrm{lcm}(\ell_1, p_2/q_2) = \mathrm{lcm}(1, p_2)$ is an integer. Similarly,

$$\ell_3 = \mathrm{lcm}\left(\ell_2, \frac{p_3}{q_3}\right) = \mathrm{lcm}(\ell_2, p_3) = \mathrm{lcm}(1, p_2, p_3)$$

is an integer and, by induction,

$$\ell_N = \mathrm{lcm}\left(\ell_{N-1}, \frac{p_N}{q_N}\right) = \mathrm{lcm}(1, p_2, \dots, p_N).$$

Now $Q = \ell_N = \mathrm{lcm}(p_2, \dots, p_N)$. Observe that $Q$ does not depend on the denominators $q_2, \dots, q_N$. It is convenient to introduce vectors $\mathbf{p} = (p_2, \dots, p_N)$ and $\mathbf{q} = (q_2, \dots, q_N)$ and write $Q(\mathbf{p})$ to highlight the dependence of $Q$ on $p_2, \dots, p_N$.

From the range constraint (18) and (20),

$$Q(\mathbf{p}) = \mathrm{lcm}(p_2, \ldots, p_N) = \frac{P}{\lambda_{\max}} \geq \frac{r_{\max}}{\lambda_{\max}} \qquad (21)$$

and from the bandwidth constraint (17) and (19),

$$\frac{\lambda_{\min}}{\lambda_{\max}} \leq \frac{p_n}{q_n} \leq 1 \qquad n = 2, \ldots, N. \qquad (22)$$

Define the objective function

$$\begin{aligned} L_2(\mathbf{p}, \mathbf{q}) &= L_1\left(\tfrac{p_2}{q_2}\lambda_{\max}, \ldots, \tfrac{p_N}{q_N}\lambda_{\max}\right) \\ &= Q^2(\mathbf{p})D(\mathbf{p},\mathbf{q}) + \frac{\gamma\lambda_{\max}^2}{D(\mathbf{p},\mathbf{q})}. \end{aligned} \qquad (23)$$

where

$$D(\mathbf{p},\mathbf{q}) = 1 + \sum_{n=2}^{N} \frac{q_n^2}{p_n^2}. \qquad (24)$$

Our optimisation problem can now be re-encoded into that of finding integers vectors

$$\hat{\mathbf{p}} = (\hat{p}_2, \ldots, \hat{p}_N), \qquad \hat{\mathbf{q}} = (\hat{q}_2, \ldots, \hat{q}_N)$$

that minimise $L_2$ subject to constraints (21) and (22). Given these minimisers, the wavelengths

$$\hat{\lambda}_n = \frac{\hat{p}_n}{\hat{q}_n} \lambda_{\max} \qquad n = 2, \ldots, N$$

are a solution of the original optimisation problem, that is, these wavelengths minimise $L_1$ subject to the bandwidth and range constraints (17) and (18). We now describe an algorithm to find minimisers $\hat{\mathbf{p}}$ and $\hat{\mathbf{q}}$.

We first discover some bounds that the minimisers $\hat{p}_n, \hat{q}_n$ must satisfy. From (22) and (24),

$$N \leq D(\hat{\mathbf{p}}, \hat{\mathbf{q}}) \leq 1 + \frac{(N-1)\lambda_{\max}^2}{\lambda_{\min}^2}. \qquad (25)$$

Also

$$Q(\hat{\mathbf{p}}) = \mathrm{lcm}(\hat{p}_2, \ldots, \hat{p}_N) \geq \hat{p}_n \qquad n = 2, \ldots, N. \qquad (26)$$

Let $\widehat{L} = L_2(\hat{\mathbf{p}}, \hat{\mathbf{q}})$ be the minimum value of $L_2$ (and also of $L_1$) and let $\widetilde{L}$ be a finite upper bound on $\widehat{L}$. For example, it suffices to choose

$$\widetilde{L} = L_1\left(\tfrac{w}{w+1}\lambda_{\max}, \ldots, \tfrac{w}{w+1}\lambda_{\max}\right) \qquad (27)$$

where $w$ is the smallest integer greater than or equal to both $r_{\max}/\lambda_{\max}$ and $\lambda_{\min}/(\lambda_{\max}-\lambda_{\min})$. With this choice $\tfrac{w}{w+1}\lambda_{\max} \in [\lambda_{\min}, \lambda_{\max}]$ so that the bandwidth constraint is satisfied and

$$\mathrm{lcm}\left(\lambda_{\max}, \tfrac{w}{w+1}\lambda_{\max}, \ldots, \tfrac{w}{w+1}\lambda_{\max}\right) = \lambda_{\max} w \geq r_{\min}$$

so that the range constraint is satisfied. Now,

$$\widetilde{L} \geq \widehat{L} = Q(\hat{\mathbf{p}})^2 D(\hat{\mathbf{p}},\hat{\mathbf{q}}) + \frac{\gamma\lambda_{\max}^2}{D(\hat{\mathbf{p}},\hat{\mathbf{q}})}$$

and using the inequalities (26) for $Q(\hat{\mathbf{p}})$ and (25) for $D(\hat{\mathbf{p}},\hat{\mathbf{q}})$ we find that,

$$\widetilde{L} \geq \widehat{L} \geq \hat{p}_n^2 N + \gamma B \qquad \text{for all } n = 2, \ldots, N$$

where

$$B = \frac{\lambda_{\min}^2 \lambda_{\max}^2}{\lambda_{\min}^2 + (N-1)\lambda_{\max}^2}.$$

Because $\hat{p}_n \geq 1$ is a positive integer we obtain the following lower and upper bounds

$$1 \leq \hat{p}_n \leq \sqrt{\frac{\widetilde{L}-\gamma B}{N}} \qquad n = 2, \ldots, N. \qquad (28)$$

Given $\hat{p}_n$ upper and lower bounds on $\hat{q}_n$ derive from the bandwidth constraint (22),

$$\hat{p}_n \leq \hat{q}_n \leq \frac{\lambda_{\max}}{\lambda_{\min}} \hat{p}_n \qquad n = 2, \ldots, N. \qquad (29)$$

To find minimisers of $L_2$, it suffices to check only those integer vectors $\hat{\mathbf{p}}$ and $\hat{\mathbf{q}}$ with elements satisfying the above two inequalities (28) and (29). Because the number of integer vectors satisfying these inequalities is finite this procedure will terminate in finite time. The number of candidate solutions that need to be checked can be reduced by incorporating the property $\gcd(\hat{p}_n, \hat{q}_n) = 1$ into the search. The number of candidates is further reduced by noting that the objective function $L_1$ is unchanged by permutation of the wavelengths $\lambda_2, \ldots, \lambda_N$. Equivalently, $L_2(\mathbf{p}, \mathbf{q})$ is unchanged if both arguments $\mathbf{p}$ and $\mathbf{q}$ undergo the same permutation. For this reason it is sufficient to suppose that the elements of $\hat{\mathbf{p}}$ are in, say, ascending order, that is, $\hat{p}_2 \leq \hat{p}_3 \leq \cdots \leq \hat{p}_N$.

Psuedocode describing the search procedure is given in Algorithm 1. The algorithm makes use of two functions psearch and qsearch that are called recursively. The integer variables $N$, $p_2, \ldots, p_N$, $q_2, \ldots, q_N$, and the real variables $\widetilde{L}, \gamma, B$ are assumed to be globally accessible to both functions psearch and qsearch. The while loop on line 1 of psearch iterates over those $p_n$ satisfying (28). The while loop on line 1 of qsearch iterates over those $q_n$ satisfying (29). The condition on line 2 of qsearch ensures that only those relatively prime $p_n, q_n$ are included in the search. Lines 6 to 8 update the minimum found value of the objective function $\widetilde{L}$ and the corresponding wavelengths whenever a new minimiser of the objective function $L_2$ is found.

This depth first search becomes computationally expensive if the number of wavelengths is not small or the minimum range $r_{\max}$ is large when compared with the maximum wavelength $\lambda_{\max}$. For this reason we now suggest some methods that accelerate the search at the expense of not necessarily guaranteeing that the true minimisers of $L$ are found. The first method simply terminates the search after a specified amount of time and takes the best wavelengths found to that point. This approach is simple, but can be highly effective because the minimisers of $L$ are regularly found well before the search completes.

The second method places a more restrictive upper bound on $\hat{p}_2, \ldots, \hat{p}_N$. The upper bound is motivated by physical constraints regularly occurring in practice that limit the accuracy to which a signal of a given wavelength can be generated [23, Sec. 5.B]. Rather than the upper bound from (28) a smaller fixed constant, say $\kappa$, is chosen and those $p_n$ satisfying $1 \leq \hat{p}_n \leq \kappa$ are searched. The condition on Line 1 of psearch is correspondingly modified to $p_n \leq \kappa$. From (29) we see that the new bound on $\hat{p}_n$ places a new bound on $\hat{q}_n$,

$$\hat{q}_n \leq \frac{\lambda_{\max}}{\lambda_{\min}} \hat{p}_n \leq \frac{\lambda_{\max}}{\lambda_{\min}} \kappa$$

Recall that the wavelengths take the form $\hat{\lambda}_n = \hat{p}_n \lambda_{\max}/\hat{q}_n$ and so this new bound limits the resolution of the wavelengths searched. Specifically, the wavelengths are restricted to the form $\lambda_{\max} p/q$ where $p \leq q$ and $q$ is smaller than $\kappa \lambda_{\max}/\lambda_{\min}$.

In practice we cannot generate sinusoidal signals with arbitrarily precise wavelengths. For example, optical interferometric experiments are limited by uncertainties in the refractive index of the medium through which the signal propagates [23, Sec. 5.B]. Audio and radio frequency devices are limited by the stability of oscillators used to generate signals. For these reasons, restricting the wavelength optimisation to a finite resolution is likely to be of little practical consequence. It might even be necessary for some applications. In practice, one might select $\kappa$ so that $\kappa \lambda_{\max}/\lambda_{\min}$ is related to the precision with which a sinusoidal signal can be generated.

---

**Algorithm 1:** Computes wavelengths optimised for the least squares range estimator.

**Input**: $N, r_{\max}, \lambda_{\min}, \lambda_{\max}, \gamma$

1 $d = \lceil \max\left(\frac{r_{\max}}{\lambda_{\max}}, \frac{\lambda_{\min}}{\lambda_{\max} - \lambda_{\min}}\right) \rceil$
2 $(\hat{\lambda}_2, \ldots, \hat{\lambda}_N) = \left(\frac{w}{w+1}\lambda_{\max}, \ldots, \frac{w}{w+1}\lambda_{\max}\right)$
3 $\widetilde{L} = L_1(\hat{\lambda}_2, \ldots, \hat{\lambda}_N)$
4 $B = \frac{\gamma \lambda_{\min}^2 \lambda_{\max}^2}{\lambda_{\min}^2 + (N-1)\lambda_{\max}^2}$
5 $p_2 = 1$
6 psearch(2)
7 **return** $(\lambda_{max}, \hat{\lambda}_2, \ldots, \hat{\lambda}_N)$

---

**Function** psearch($n$)

**Input**: $n \in \{2, \ldots, N\}$

1 **while** $p_n^2 \leq (\widetilde{L} - \gamma B)/N$ **do**
2     $q_n = p_n$
3     qsearch($n$)
4     $p_n = p_n + 1$

---

**Function** qsearch($n$)

**Input**: $n \in \{2, \ldots, N\}$

1 **while** $q_n \leq p_n \lambda_{max}/\lambda_{min}$ **do**
2     **if** $\gcd(p_n, q_n) = 1$ **then**
3        **if** $n < N$ **then**
4           $p_{n+1} = p_n$
5           psearch($n+1$)
6        **else if** $L_2(\mathbf{p}, \mathbf{q}) \leq \widetilde{L}$ and $\text{lcm}(\mathbf{p}) \geq \frac{r_{max}}{\lambda_{max}}$ **then**
7           $\widetilde{L} = L_2(\mathbf{p}, \mathbf{q})$
8           $\hat{\lambda}_n = \lambda_{\max} p_n/q_n \quad n = 2, \ldots, N$
9     $q_n = q_n + 1$

## VI. SIMULATION RESULTS

We present the results of Monte-Carlo simulations with the least squares range estimator, the excess fractions estimator [20], the algebraic method of Falaggis et al. [23], and range estimators based on the single-stage and multi-stage CRT algorithms of Xiao et al. [31]. In each simulation the phase noise variables $\Phi_1, \ldots, \Phi_N$ are wrapped normally distributed, that is, $\Phi_n = \langle \epsilon_n \rangle$ where $\epsilon_1, \ldots, \epsilon_N$ are independent and normally distributed with zero mean and variance $\sigma^2$. The number of Monte-Carlo trials used for each value of $\sigma^2$ is $10^5$.

Figure 2 shows the sample mean square error of these estimators for $N = 3$ wavelengths. In each simulation the true range is $r_0 = 6\pi$ and we consider two different sets of wavelengths

$$A = \{2, 3, 5\}, \qquad B = \{\tfrac{30}{13}, \tfrac{15}{4}, 5\}.$$

The wavelengths from both sets are contained in the interval $[2, 5]$ and the identifiable range is $\text{lcm}(A) = \text{lcm}(B) = 30$. The wavelengths $A$ are used in the simulations of Li et. al. [33] and $B$ are optimised wavelengths given by Algorithm 1. When the noise variance is small the probability that the wrapping variables are correct is large and so we expect the mean square error to be similar to (14). This predicted mean square error is shown by the solid line for wavelengths $A$ and by the dashed line for wavelengths $B$. Observed from Figure 2 that these predictions accurately model the behaviour of the least squares estimator when $\sigma^2$ is small.

Wavelengths $A$ result in slightly reduced sample mean square error compared with $B$ when $\sigma^2$ is small. As $\sigma^2$ increases the sample mean square error exhibits a 'threshold' effect and increases suddenly. The threshold occurs at $\sigma^2 \approx 5 \times 10^{-4}$ with wavelengths $A$ and $\sigma^2 \approx 9 \times 10^{-4}$ with wavelength $B$ for the least squares estimator. Wavelengths $B$ are more accurate than $A$ when $\sigma^2$ is greater than approximately $5 \times 10^{-4}$. The threshold for the CRT and excess fractions based range estimators occurs at approximately the same value of $\sigma^2$ with wavelengths $A$. The CRT estimator performs poorly with wavelengths $B$. The threshold for the excess fractions estimator is similar to that of the least squares estimator with wavelength $B$. However, the mean square error of the excess fraction estimator is larger than that of the least squares estimator when the noise variance is small.

Figure 3 displays the simulations results when there are $N = 4$ wavelengths from the sets

$$C = \{\tfrac{101039}{66}, \tfrac{1076285}{682}, \tfrac{198036440}{125389}, \tfrac{17572}{11}\},$$

$$D = \{1528, \tfrac{3868970284693}{25 \times 10^8}, \tfrac{156953786407767}{10^{11}}, \tfrac{17572}{11}\}.$$

The wavelengths from both sets are contained in the interval $[1528, \tfrac{17572}{11}]$. Wavelengths $D$ are those selected for the excess fractions estimator using the procedure described in [21]. These wavelengths are measured in nanometers in [21]. The least common multiple of $D$ is greater than $2 \times 10^{22}$ meters. However, the maximum range of the excess fractions estimator is not the least common multiple, but is instead what is called the *unambiguous measurement range* (UMR) [21] and is $1.8 \times 10^7 \text{nm} = 0.018 \text{m}$ in this case. Wavelengths $C$ are optimised for the least squares estimator using Algorithm 1 with $r_{\max} = 1.8 \times 10^7$ equal to the UMR. To speed up the search we put $\kappa = 1000$ as described in Section V. The

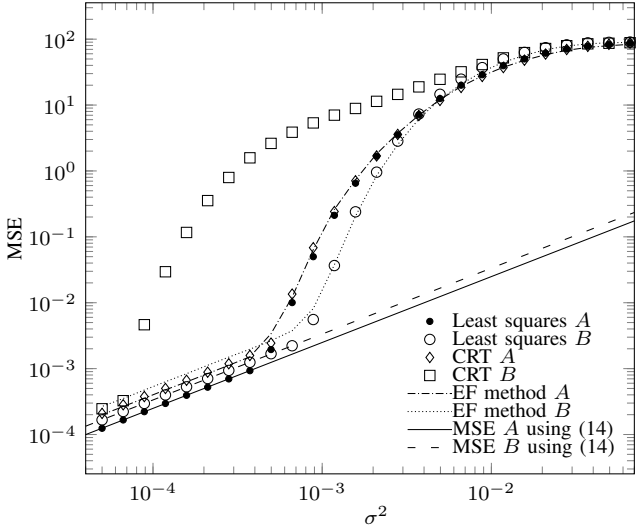

Fig. 2. Sample mean square error of the least squares range estimator, the excess fraction based range estimator [20] and the range estimator based on the single stage and multi-stage CRT algorithms of Xiao et. al. [31] with $N = 3$ wavelengths.

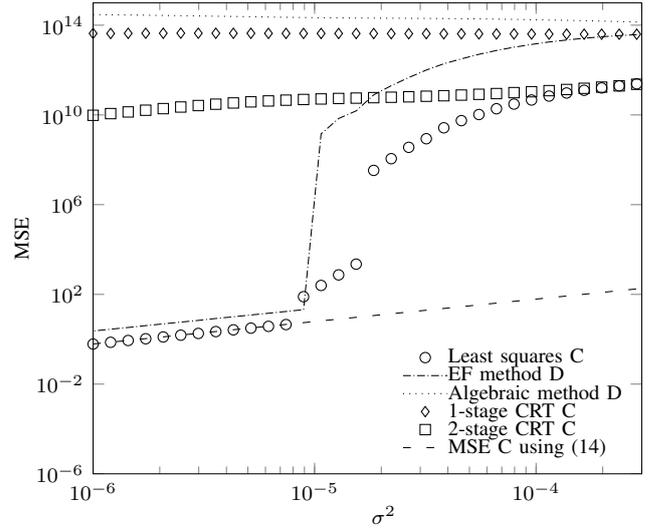

Fig. 3. Sample mean square error of the least squares range estimator, the excess fraction based range estimator [20], the algebraic method of Falaggis et al. [23] and the range estimator based on the single stage and multi-stage CRT algorithms of Xiao et. al. [31] with $N = 4$ wavelengths.

identifiable range with wavelengths $C$ is

$$\text{lcm}(C) = 198036440/11 \approx 18003312 > 1.8 \times 10^7.$$

In each simulation the true range $r_0 = 4000000\pi \approx 0.7 r_{\max}$. It can be observed from this figure that the single and multi-stage CRT estimators [31] and the algebraic method of Falaggis et al. [23] perform very poorly when compared with the excess fractions [20] and the least squares estimator. When $\sigma^2$ is less than $\approx 8 \times 10^{-6}$ the least squares estimator is slightly more accurate than the excess fractions estimator. The thresholds for the excess fractions and the least squares estimators occur at approximately $8 \times 10^{-6}$ and $1.5 \times 10^{-5}$ respectively. The least squares estimator is the most accurate among the estimators. It can also be observed that (14) provides a very good approximation for the MSE of the least squares estimator.

In another simulation in Figure 4 we compare the sample mean square error of the least squares range estimator, the excess fractions [20] and the single stage and multi-stage CRT based estimators of Xiao et. al. [31] with $N = 5$ wavelengths. In each simulation the true range $r_0 = 300\pi$. Two different sets of wavelengths are considered,

$$E = \{2, 3, 5, 7, 11\}, \quad F = \{\tfrac{22}{3}, \tfrac{66}{17}, \tfrac{77}{18}, \tfrac{110}{31}, 11\}.$$

The wavelengths from both sets are contained in the interval $[2, 11]$ and $P = \text{lcm}(E) = \text{lcm}(F) = 2310$ so that the identifiable range is the same. Wavelengths $E$ are relatively prime integers and are used in [33]. Wavelengths $F$ are obtained using Algorithm 1 with $\kappa = 15$.

The behaviour of the least squares, excess fractions and single-stage CRT estimators is similar for the wavelengths $E$. No benefit is gained by applying the multi-stage CRT estimator with wavelengths $E$. When the noise variance $\sigma^2$ is small the least squares estimator exhibits slightly smaller mean square error than the excess fractions and CRT estimators. The threshold for all of the estimators occurs at $\sigma^2 \approx 8 \times 10^{-5}$ with wavelengths $E$. Different behaviour is exhibited with wavelength $F$. When the noise variance $\sigma^2$ is small the least squares estimator exhibits slightly smaller mean square error with wavelengths $E$ than with $F$. However, the threshold with wavelengths $F$ occurs at $\sigma^2 \approx 2 \times 10^{-4}$. Wavelengths $F$ are more accurate than $E$ when $\sigma^2$ is greater than approximately $8 \times 10^{-5}$. The single-stage CRT estimator performs comparatively poorly with wavelengths $F$. A small improvement is gained by use of the multi-stage CRT estimator by splitting the wavelengths from $F$ into two sets $\{\tfrac{110}{31}, 11\}$ and $\{\tfrac{22}{3}, \tfrac{66}{17}, \tfrac{77}{18}\}$. Simulations indicate that this is the best splitting of the wavelengths for the multi-stage CRT estimator in this case. The excess fractions estimator performs very poorly with wavelengths $F$.

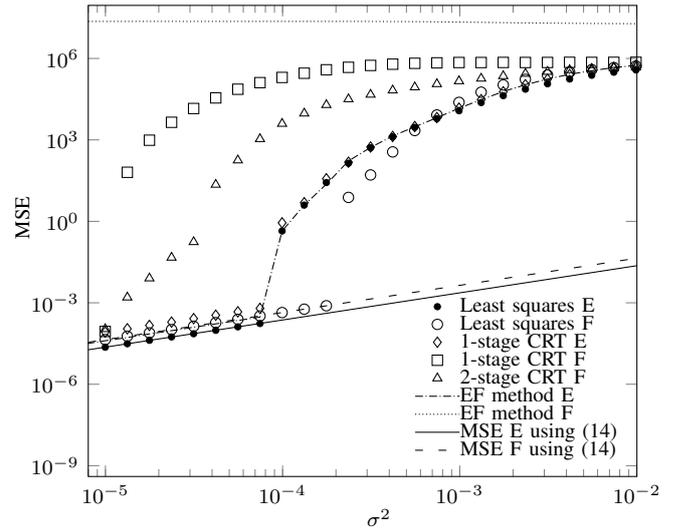

Fig. 4. Sample mean square error of the least squares range estimator, the excess fraction based range estimator [20] and the range estimator based on the single stage and multi-stage CRT algorithms of Xiao et. al. [31] with $N = 5$ wavelengths.

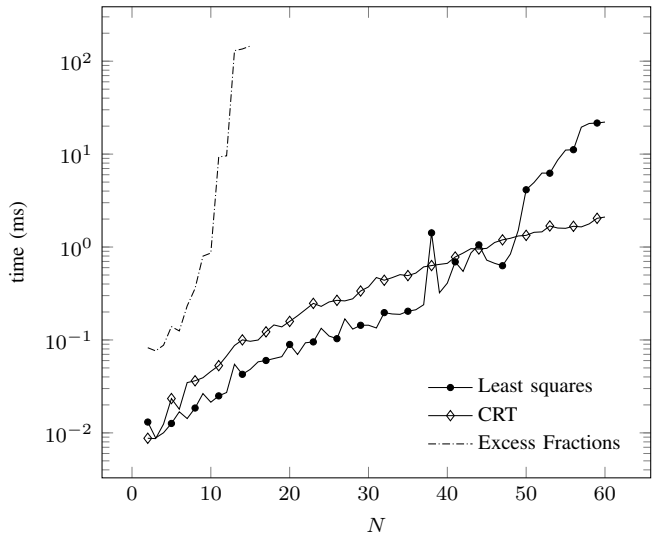

Fig. 5. Computation time benchmark: Comparison of the least squares estimator, the CRT estimator and the excess fractions estimator.

Figure 5 shows the computation time required for the least squares estimator computed using a sphere decoder (Section III), the single stage CRT estimator of Xiao et. al. [31], and the excess fractions based estimator [20] as the number of wavelengths $N$ increases. The wavelengths are set to integers $\{1, 2, 3, \ldots, N\}$ in each benchmark. In these benchmarks the least squares estimator is faster than the CRT for $N$ less than 38. However, for large $N$ the least square estimator computed using the sphere decoder becomes prohibitively expensive. The excess fractions estimator is computationally expensive even for a small number of wavelengths. The computational complexity of the excess fractions estimator increases with the ratio between the unambiguous measurement range (UMR) and the smallest wavelength. The complexity can be prohibitive even for three wavelengths if this ratio is large.

## VII. CONCLUSION

We have considered the problem of selecting an optimised set of wavelengths for the least squares range estimator. Using some interesting properties of lattices we have formulated an optimisation criterion that aims to minimise the mean square error of the estimator. Based on this optimisation criterion a depth first search algorithm is developed that outputs a set of wavelengths that typically yield smaller mean square error when employed with the least squares estimator. Simulations indicate that the wavelengths obtained using this algorithm outperform the existing wavelength selection methods for the excess fractions range estimator and also outperform the CRT based range estimators.

Our optimisation criterion is based upon the upper bound (12) on the probability probability that an $m$-variate normal random lies inside the Voronoi cell of a lattice. It would be interesting to investigate how the lower bound (13) might be used to produce alternative wavelength optimisation procedures.

## REFERENCES

[1] E. Jacobs and E. W. Ralston, "Ambiguity resolution in interferometry," *IEEE Trans. Aerospace Elec. Systems*, vol. 17, no. 6, pp. 766–780, Nov. 1981.
[2] J. M. M. Anderson, J. M. Anderson, and E. M. Mikhail, *Surveying, theory and practice*, WCB/McGraw-Hill, 1998.
[3] P. J. G. Teunissen, "The LAMBDA method for the GNSS compass," *Artificial Satellites*, vol. 41, no. 3, pp. 89–103, 2006.
[4] P. J. G. Teunissen, "The least-squares ambiguity decorrelation adjustment: a method for fast GPS integer ambiguity estimation," *Journal of Geodesy*, vol. 70, pp. 65–82, 1995.
[5] M. Frank, M. Plaue, H. Rapp, U. Koethe, B. Jhne, and F. A. Hamprecht, "Theoretical and experimental error analysis of continuous-wave time-of-flight range cameras," *Optical Engineering*, vol. 48, no. 1, Jan. 2009.
[6] A. Benedetti, "Methods and systems for geometric phase unwrapping in time of flight systems," United States patent, 0049767 A1, Feb. 2014.
[7] S. D. Chitte, S. Dasgupta, and D. Zhi, "Distance estimation from received signal strength under log-normal shadowing: Bias and variance," *IEEE Signal Process. Letters*, vol. 16, no. 3, pp. 216–218, March 2009.
[8] H.-C. So and L. Lin, "Linear least squares approach for accurate received signal strength based source localization.," *IEEE Trans. Sig. Process.*, vol. 59, no. 8, pp. 4035–4040, 2011.
[9] X. Li and K. Pahlavan, "Super-resolution TOA estimation with diversity for indoor geolocation," *IEEE Trans. Wireless Commun.*, vol. 3, no. 1, pp. 224–234, Jan 2004.
[10] S. Lanzisera, D. Zats, and K.S.J. Pister, "Radio frequency time-of-flight distance measurement for low-cost wireless sensor localization," *IEEE Sensors Journal*, vol. 11, no. 3, pp. 837–845, March 2011.
[11] Fauzia Ahmad, Moeness G. Amin, and Paul D. Zemany, "Performance analysis of dual-frequency CW radars for motion detection and ranging in urban sensing applications," *Proc. SPIE 6547, Radar Sensor Technology XI*, vol. 6547, pp. 65470K–65470K–8, May 2007.
[12] A. Povalac and J. Sebesta, "Phase difference of arrival distance estimation for RFID tags in frequency domain," in *RFID-Technologies and Applications (RFID-TA), 2011 IEEE International Conference on*, Sept 2011, pp. 188–193.
[13] K. Thangarajah, R. Rashizadeh, S. Erfani, and M. Ahmadi, "A hybrid algorithm for range estimation in RFID systems," in *Electronics, Circuits and Systems (ICECS), 2012 19th IEEE International Conference on*, Dec 2012, pp. 921–924.
[14] Burkhard Stiller, Thomas Bocek, Fabio Hecht, Guilherme Machado, Peter Racz, and Martin Waldburger, "Real-Time Kinematic Surveying - Training Guide," Tech. Rep., Trimble Navigation Limited, 01 2003.
[15] D. Grejner-Brzezinska, I. Kashani, P. Wielgosz, D. Smith, P. Spencer, D. Robertson, and G. Mader, "Efficiency and reliability of ambiguity resolution in network-based real-time kinematic GPS," *J. of Surveying Eng.*, vol. 133, no. 2, pp. 56–65, 2007.
[16] D. Odijk, P. Teunissen, and B. Zhang, "Single-frequency integer ambiguity resolution enabled GPS precise point positioning," *J. of Surveying Eng.*, vol. 138, no. 4, pp. 193–202, 2012.
[17] N. I. Fisher, *Statistical analysis of circular data*, Cambridge University Press, 1993.
[18] C. E. Towers, D. P. Towers, and J. D. C. Jones, "Optimum frequency selection in multifrequency interferometry," *Optics Letters*, vol. 28, pp. 887–889, 2003.
[19] Catherine E. Towers, David P. Towers, and Julian D. C. Jones, "Generalized frequency selection in multifrequency interferometry," *Opt. Lett.*, vol. 29, no. 12, pp. 1348–1350, Jun 2004.

[20] Konstantinos Falaggis, David P. Towers, and Catherine E. Towers, "Method of excess fractions with application to absolute distance metrology: theoretical analysis," *Appl. Opt.*, vol. 50, no. 28, pp. 5484–5498, Oct 2011.

[21] K. Falaggis, David P. Towers, and Catherine E. Towers, "Method of excess fractions with application to absolute distance metrology: Wavelength selection and the effects of common error sources," *Appl. Opt.*, vol. 51, no. 27, pp. 6471–6479, Sep 2012.

[22] Konstantinos Falaggis, David P. Towers, and Catherine E. Towers, "Method of excess fractions with application to absolute distance metrology: analytical solution," *Appl. Opt.*, vol. 52, no. 23, pp. 5758–5765, Aug 2013.

[23] Konstantinos Falaggis, David P. Towers, and Catherine E. Towers, "Algebraic solution for phase unwrapping problems in multiwavelength interferometry," *Appl. Opt.*, vol. 53, no. 17, pp. 3737–3747, Jun 2014.

[24] O. Ore, "The general Chinese remainder theorem," *The American Mathematical Monthly*, vol. 59, pp. 365–370, 1952.

[25] O. Goldreich, D. Ron, and M. Sudan, "Chinese remaindering with errors," *IEEE Trans. Inform. Theory*, vol. 46, no. 4, pp. 1330–1338, Jul 2000.

[26] X. G. Xia and K. Liu, "A generalized Chinese Remainder Theorem for residue sets with errors and its application in frequency determination from multiple sensors with low sampling rates," *IEEE Signal Process. Letters*, vol. 12, no. 11, pp. 768–771, Nov 2005.

[27] X. G. Xia and G. Y. Wang, "Phase unwrapping and a robust Chinese Remainder Theorem," *IEEE Signal Process. Letters*, vol. 14, no. 4, pp. 247–250, April 2007.

[28] X. W. Li and X. G. Xia, "A fast robust Chinese Remainder Theorem based phase unwrapping algorithm," *IEEE Signal Process. Letters*, vol. 15, pp. 665–668, 2008.

[29] W. Wang and X. G. Xia, "A closed-form robust Chinese Remainder Theorem and its performance analysis," *IEEE Trans. Sig. Process.*, vol. 58, no. 11, pp. 5655–5666, Nov 2010.

[30] B. Yang, W. Wang, X. G. Xia, and Q. Y. Yin, "Phase detection based range estimation with a dual-band robust Chinese Remainder Theorem," *Science China Information Sciences*, vol. 57, no. 2, pp. 1–9, 2014.

[31] Li Xiao, Xiang-Gen Xia, and Wenjie Wang, "Multi-stage robust Chinese Remainder Theorem," *IEEE Trans. Sig. Process.*, vol. 62, no. 18, pp. 4772–4785, Sept 2014.

[32] A. Hassibi and S. P. Boyd, "Integer parameter estimation in linear models with applications to GPS," *IEEE Trans. Sig. Process.*, vol. 46, no. 11, pp. 2938–2952, Nov 1998.

[33] W. Li, X. Wang, X. Wang, and B. Moran, "Distance estimation using wrapped phase measurements in noise," *IEEE Trans. Sig. Process.*, vol. 61, no. 7, pp. 1676–1688, 2013.

[34] A. Akhlaq, R. McKilliam, and R. Subramanian, "Basis construction for range estimation by phase unwrapping," *Signal Processing Letters, IEEE*, vol. PP, no. 99, pp. 1–1, 2015.

[35] E. Agrell, T. Eriksson, A. Vardy, and K. Zeger, "Closest point search in lattices," *IEEE Trans. Inform. Theory*, vol. 48, no. 8, pp. 2201–2214, Aug. 2002.

[36] J. H. Conway and N. J. A. Sloane, *Sphere packings, lattices and groups*, Springer, New York, 3rd edition, 1998.

[37] J. Martinet, *Perfect lattices in Euclidean spaces*, Springer, 2003.

[38] H. Cohen, *A course in computational algebraic number theory*, Springer Verlag, 1993.

[39] R. G. McKilliam, *Lattice theory, circular statistics and polynomial phase signals*, Ph.D. thesis, University of Queensland, Australia, December 2010.

[40] A. Pollok A. Akhlaq, R. G. McKilliam, "Robustness of the least squares range estimator," accepted to the Australian Communications Theory Workshop (AusCTW), 2016.

[41] K. V. Mardia and P. Jupp, *Directional Statistics*, John Wiley & Sons, 2nd edition, 2000.

[42] D. Micciancio and P. Voulgaris, "A deterministic single exponential time algorithm for most lattice problems based on Voronoi cell computations," *SIAM Journal on Computing*, vol. 42, no. 3, pp. 1364–1391, 2013.

[43] R. McKilliam, A. Grant, and I. Clarkson, "Finding a closest point in a lattice of Voronoi's first kind," *SIAM Journal on Discrete Mathematics*, vol. 28, no. 3, pp. 1405–1422, 2014.

[44] J. H. Conway and N. J. A. Sloane, "A fast encoding method for lattice codes and quantizers," *IEEE Trans. Inform. Theory*, vol. 29, no. 6, pp. 820–824, Nov 1983.

[45] U. Erez and R. Zamir, "Achieving $1/2 \log(1 + SNR)$ on the AWGN channel with lattice encoding and decoding," *IEEE Trans. Inform. Theory*, vol. 50, no. 10, pp. 2293–2314, Oct. 2004.

[46] I. V. L. Clarkson, "Approximate maximum-likelihood period estimation from sparse, noisy timing data," *IEEE Trans. Sig. Process.*, vol. 56, no. 5, pp. 1779–1787, May 2008.

[47] R. G. McKilliam and I. V. L. Clarkson, "Identifiability and aliasing in polynomial-phase signals," *IEEE Trans. Sig. Process.*, vol. 57, no. 11, pp. 4554–4557, Nov. 2009.

[48] R. G. McKilliam, B. G. Quinn, I. V. L. Clarkson, and B. Moran, "Frequency estimation by phase unwrapping," *IEEE Trans. Sig. Process.*, vol. 58, no. 6, pp. 2953–2963, June 2010.

[49] R. G. McKilliam, B. G. Quinn, and I. V. L. Clarkson, "Direction estimation by minimum squared arc length," *IEEE Trans. Sig. Process.*, vol. 60, no. 5, pp. 2115–2124, May 2012.

[50] D. Micciancio and O. Regev, "Lattice based cryptography," in *Post Quantum Cryptography*, D .J. Bernstein, J. Buchmann, and E. Dahmen, Eds. Springer, 2009.

[51] R. G. McKilliam, B. G. Quinn, I. V. L. Clarkson, B. Moran, and B. N. Vellambi, "Polynomial phase estimation by least squares phase unwrapping," *IEEE Transactions on Signal Processing*, vol. 62, no. 8, pp. 1962–1975, April 2014.

[52] R.G. McKilliam, I.V.L. Clarkson, and B.G. Quinn, "Fast sparse period estimation," *IEEE Signal Processing Letters*, vol. 22, no. 1, pp. 62–66, Jan 2015.